# Electrohydrodynamics of compound droplet in a microfluidic confinement

Somnath Santra[1], Sayan Das[1] and Suman Chakraborty[1,†]

[1]*Department of Mechanical Engineering, Indian Institute of Technology Kharagpur, Kharagpur, West Bengal - 721302, India*

The present study deals with the dynamics of a double emulsion confined in a microchannel under the presence of a uniform electric field. Towards investigating the non-trivial electrohydrodynamics of a compound droplet, confined in between two parallel plate electrodes, a phase field approach has been adopted. Under the assumption of negligible fluid inertia and small shape deformation, an asymptotic model is also developed to predict the transient as well as the steady state droplet dynamics in the limiting case of an unbounded suspending medium. The phases involved are either assumed to be perfect dielectric or leaky dielectric. Subsequent investigation shows that shape deformation of either of the interfaces of the droplet increase with rise in the channel confinement for a perfect dielectric system. However, for a leaky dielectric system, the deformation of inner droplet and outer droplet can increase or decrease with the rise in channel confinement depending on the electric properties (conductivity and permittivity) of the system. It is also observed that the presence of inner droplet for a compound droplet suppresses the breakup of the outer droplet. The present study further takes into account the effect of eccentricity of the inner droplet. Depending on the relative magnitude of the permittivity and conductivity of the system, the inner droplet, if eccentrically located, may exhibit a to and fro or a simple translational motion. It is seen that increase in the channel confinement results in a translational motion of the inner droplet thus rendering its motion independent of any electrical properties. Since the compound droplet model approximately mimics the structure of a cell, the current investigation is expected to provide a significant impact towards predicting the effect of an externally imposed electrical field on the dynamics of a cell in any fluidic confinement and hence has the potential towards widespread applications in the field of medical diagnostics.

† E-mail address for correspondence: suman@mech.iitkgp.ernet.in

# 1. Introduction

The fundamental understanding of the deformation of a droplet subjected to uniform electric field is of immense interest and currently holds the attention of different scientific and industrial communities because of its wide range of applications from natural to modern day microfluidic processes (Macky 1931; C.G Garton and Z. Krasucki 1964; Anna 2016; Teh et al. 2008; Zhu & Fang 2013; Mhatre et al. 2015). Some of its applications are directed towards the enhancement of coalescence and demixing in emulsion processes (Ptasinski & Kerkhof 1992), ink-jet printing (Basaran 2002; Chen et al. 2003), as well as increase in the efficiency of combustion processes (Tao et al. 2008). In bio-medical fields too, droplets of tiny volume are sometimes used as micro reactors (Tsutsui & Ho 2009; Zhou et al. 2005). While we have found a ubiquitous appearance of single phase droplet in different flow processes, encapsulated droplet (or a compound droplet) also has significant attribution in a growing number of new applications. Some of its earlier well known applications state that compound droplets serve the purpose of liquid membranes in selective mass transfer process (T. Araki and H. Tsukube 1990), in separation of hydrocarbons (Li 1971), blood oxygenation (LI & Asher 1973), purifications of water (Li & Shrier 1972) and controlled release of drugs (Mataumoto & Kang 1989). In the recent times, the focus of increasing attention is directed towards biomedical applications that include distortion as well as recovery of white blood cells in presence of incipient flow (Kan et al. 1998; Tasoglu et al. 2010). These studies use the compound droplet model to mimic the dynamics of a leukocyte, where the core (inner droplet) and shell (outer droplet) denote the cell nucleus and the cytoplasm, respectively.

In all the above applications involving a compound droplet, electric field can be used as an effective means to enhance the efficiency of the process. However, for a successful implementation of the same, an in depth knowledge of dynamics of liquid-liquid interface in presence of electric field is very crucial. From the pioneering work of Taylor (1966), it has been revealed that the interfacial electric stress generated due to the presence of electric field not only deforms the interface into a prolate or oblate shape, but also creates a hydrodynamic flow in and around the droplet. For denoting the sense of deformation, Taylor (1966) has analytically derived a discriminating function ($\Omega_T$) in terms of $S$, $R$ and $\lambda$, where $S$, $R$ and $\lambda$ represent the ratios of the permittivity, conductivity and viscosity of the droplet and the surrounding phase, respectively. For $\Omega_T < 0$ and $R < S$, the direction of flow circulation is from poles (aligned in the electric field direction) to equator and the droplet deforms into oblate shape. On the other side, the flow direction reverses and the prolate deformation of the droplet is observed for $R > S$ and $\Omega_T > 0$. Later, several studies (Das & Saintillan 2017; Feng & Scott 2006; Feng 1999; Lac & Homsy 2007; Salipante & Vlahovska 2010; Nganguia et al. 2016) have been performed on the EHD of single droplet that explored the different concepts towards a better understanding of physics involved.

Unlike the case of single droplets, significant attention have not been given towards the electrohydrodynamics of compound droplets. The earliest study was performed by Gouz and Sadhal (1989), where they have looked into the setting of a compound droplet in the presence of gravity. They investigated the stability of the compound droplet in the presence of an electric field and suggested that a stable equilibrium position of the droplet can be achieved even in the presence of gravitational force. Tsukada et al. (1997) have numerically shown that the intensity of electric field and ratio of core phase to shell phase volume have a significant effect on the orientation of fluid circulation as well as the interface deformation. Later on, Ha and Yang (1999) have studied this problem analytically and found out the possibilities of the formation of double structured vortex. In recent years, Behjatian & Esmaeeli (2015) have looked into the EHD of the droplets in details and shown that there are four types of flow pattern depending on the direction of external fluid flow and the number of vortices. They have also pointed out four possible modes of droplet deformation on the basis of relative strength of normal electric and hydrodynamic stress. In a related study, Soni et al. (2013) have shown that the presence of inner droplet alters the deformation of outer droplet.

Till date, several studies have been performed that has thoroughly investigated the influence of an externally imposed electric field on the dynamics of a liquid column (Reddy & Esmaeeli 2009; Behjatian & Esmaeeli 2013a; Behjatian & Esmaeeli 2014; Rhodes et al. 1989; Esmaeeli 2016). However, the deformation dynamics of a compound liquid column (or a two-dimensional compound droplet) in the presence of uniform electric field is barely explored. The liquid column model is a close approximation to that of a 2D droplet and yet sheds light on the important physics involved in its behavior in the presence of an electric field, which facilitates us with a better understanding of the underlying concepts. In a recent study, Behjatian and Esmaeeli (2013b) have studied the steady state flow pattern and the modes of deformation of a co-axial liquid column in the presence of a transverse electric field. They have pointed out four different modes of steady state flow pattern and deformation similar to a compound droplet. In a subsequent study (Behjatian & Esmaeeli 2015), they have incorporated the transient electrohydrodynamics in their analysis.

Though the general features of steady state and transient electrohydronamics of droplet have been studied well, nonetheless the EHD of a compound liquid column in the presence of a channel confinement is yet unexplored. In two consecutive studies (Esmaeeli & Behjatian 2012; Behjatian & Esmaeeli 2013a), Behjatian and Esmaeelli have studied the EHD of single liquid column (and single droplet) in a cylindrical (and spherical) confined domain. Their studies depict that wall confinement can increase or decrease the magnitude of electric field strength as well as the fluid flow intensity depending on the value of $R$. Furthermore, they have also shown that the steady state shape of the droplet changes from oblate to prolate with increase in the channel confinement. Recently, Santra et al. (2018) studied the effect of channel confinement on the deformation as well as break up of a single droplet subjected to a uniform electric field. In the present study, we have analyzed the effect of channel confinement on the steady state morphology of a compound droplet, transient evolution of its configuration as well as the motion of an eccentric inner droplet with respect to the outer. These provide us with a fundamental understanding of the functionalities of a wide range of microfluidic devices. Present analysis highlights that the deformation of either of the interfaces can change depending on the degree of confinement of the system similar to that of a single droplet. Furthermore, it is also shown that an eccentrically placed inner droplet, following a to and fro motion in weakly confined domain undergoes linear translational motion in a confined domain. To support our numerical findings, an asymptotic analysis is also performed under the assumption of small shape deformation.

## 2. Problem formulation

### 2.1 System description

Figure 1 shows the schematic of the physical system, where a neutrally buoyant compound droplet is suspended in another medium in between two parallel plates under the influence of uniform electric field. All the phases are Newtonian, immiscible and incompressible with the radius of the outer and the inner undeformed interfaces being $a_1$ and $a_2$ respectively. Properties such as viscosity, electrical permittivity, electrical conductivity and surface tension are represented by $\mu$, $\varepsilon$, $\sigma$, and $\gamma$ respectively. Subscript 1, 2 and 3 denote the inner phase (inner droplet, fluid 1), outer phase (outer droplet, fluid 2) and carrier fluid phase (fluid 3) respectively whereas the double subscript 12 and 23 represents the inner and outer interface respectively. For the numerical analysis, the inner droplet is taken to be located eccentrically with respect to the outer droplet, the eccentricity being $e$ (refer to figure 1). The separating distance between the two wall is $2H$ and the electric field strength is denoted by $\bar{E}_\infty$. Thus the electric potentials of top wall and bottom wall electrodes are taken as $\bar{H}\bar{E}_\infty$ and $-\bar{H}\bar{E}_\infty$ respectively.

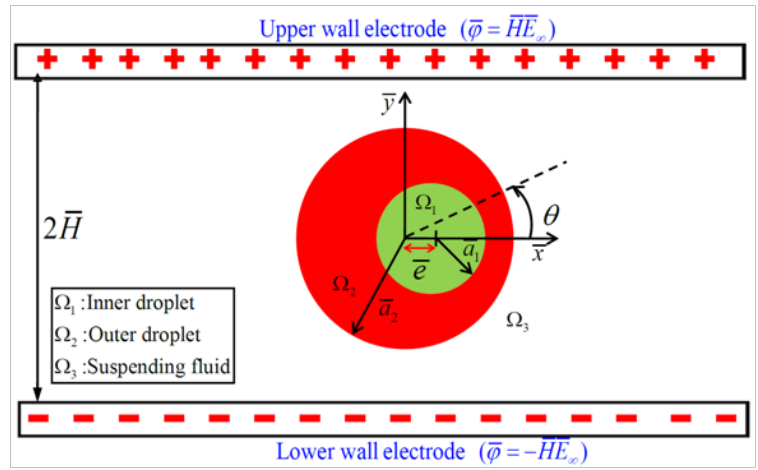

FIGURE 1. Schematic of the problem setup displaying a compound droplet placed in between two parallel plate electrodes. The radius of the core (inner droplet) and shell (outer droplet) are $\bar{a}_1$ and $\bar{a}_2$ respectively.

The present analysis utilizes a two-dimensional Cartesian coordinate system ($\bar{x}$-$\bar{y}$) that is attached at the droplet centroid.

## 2.2. Governing equations and boundary conditions

In the current study, we have taken leaky dielectric model into consideration. In leaky dielectric model, the free charges are assumed to instantaneously move to the interface such the bulk remains electrically neutral. Thus, the equations governing the electric field and the flow field are decoupled from each other and the problem is reduced to an electrostatic problem. Furthermore, the applied electric field is also irrotational and it is expressed as $\bar{\mathbf{E}} = -\bar{\nabla}\bar{\varphi}$ (Reddy & Esmaeeli 2009). Therefore, the equation governing the electric potential ($\bar{\varphi}$) for the $i$th fluid ($i = 1, 2, 3$) is represented in the following shape

$$\bar{\nabla}^2 \bar{\varphi}_i = 0. \tag{1}$$

Now the electric potential distribution satisfies the following boundary conditions:

(i) inside the inner droplet, $\bar{\varphi}_1$ is finite. Outside the outer droplet, $\bar{\varphi}_3$ satisfies the following boundary conditions

$$\left. \begin{array}{l} \text{at } \bar{y} = \bar{H},\ \bar{\varphi}_3 = \bar{H}\bar{E}_\infty, \\ \text{at } \bar{y} = -\bar{H},\ \bar{\varphi}_3 = -\bar{H}\bar{E}_\infty, \end{array} \right\} \tag{2}$$

(ii) at the deformed interfaces ($\Pi_{ij}$), $\bar{\varphi}_i$ is continuous and expressed as

$$\left[\bar{\varphi}_i\right]_{\Pi_{ij}} = \left[\bar{\varphi}_e\right]_{\Pi_{ij}} \quad \text{at } \bar{r} = \bar{r}_{\Pi_{ij}}(\theta, \bar{t}), \tag{3}$$

where $\bar{r} = \bar{r}_{\Pi_{ij}}(\theta, \bar{t})$ denotes the radial location of the deformed interface.

(iii) at the deformed interfaces ($\Pi_{ij}$), the normal component of electrical current density is continuous and represented in the following format

$$\left[\sigma_i \bar{\nabla}\bar{\varphi}_i \cdot \mathbf{n}_{ij}\right]_{\Pi_{ij}} = \left[\sigma_j \bar{\nabla}\bar{\varphi}_j \cdot \mathbf{n}_{ij}\right]_{\Pi_{ij}} \quad \text{at } \bar{r} = \bar{r}_{\Pi_{ij}}(\theta, \bar{t}), \tag{4}$$

where the normal unit vector at the interface $\Pi_{ij}$ in the outward direction is denoted by $\mathbf{n}_{ij}$ (Mandal, Bandopadhyay, et al. 2016). We have employed general notation $[\ ]_{\Pi_{ij}}$ for denoting any quantity at the interface $\Pi_{ij}$.

The velocity ($\bar{\mathbf{u}}$) and pressure field ($\bar{p}$) for $i$th fluid ($i = 1, 2, 3$) are obtained by solving the continuity and Navier-Stokes equation and can be expressed as

$$\overline{\nabla}\cdot\overline{\mathbf{u}}_i = 0, \quad \rho\left(\frac{\partial \overline{\mathbf{u}}_i}{\partial \overline{t}} + \overline{\nabla}\cdot(\overline{\mathbf{u}}_i\overline{\mathbf{u}}_i)\right) = -\overline{\nabla}\overline{p}_i + \overline{\nabla}\cdot\left[\mu_i\left\{\overline{\nabla}\overline{\mathbf{u}}_i + \left(\overline{\nabla}\overline{\mathbf{u}}_i\right)^T\right\}\right]. \tag{5}$$

The absence of any body force term in above equation can be clearly noted. The reason for the same can be attributed to the charge free bulk medium with constant permittivity. However, the coupling between the electrostatic and hydrodynamic flow field is present at the interface through stress balance which is apparent from equation (7) and is stated below. At the centroid of the inner droplet (phase 1), the velocity field is finite, however it vanishes at the electrode walls. At the interfaces, the velocity field fulfills the no slip as well as the no penetration boundary condition, which can be expressed as follows

$$\left. \begin{array}{l} [\overline{\mathbf{u}}_i]_{\Pi_{ij}} = [\overline{\mathbf{u}}_j]_{\Pi_{ij}} \quad \text{at } \overline{r} = \overline{r}_{\Pi_{ij}}(\theta,\overline{t}), \\[2mm] [\overline{\mathbf{u}}_i \cdot \mathbf{n}_{ij}]_{\Pi_{ij}} = [\overline{\mathbf{u}}_j \cdot \mathbf{n}_{ij}]_{\Pi_{ij}} = \dfrac{d\overline{\mathbf{r}}_{\Pi_{ij}}}{d\overline{t}} \quad \text{at } \overline{r} = \overline{r}_{\Pi_{ij}}(\theta,\overline{t}), \end{array} \right\} \tag{6}$$

where, $i, j \in (1,2,3)$. The stress balance condition can be represented in the following way,

$$\left[\left(\overline{\tau}_j^H + \overline{\tau}_j^E\right)\cdot\mathbf{n}_{ij}\right]_{\Pi_{ij}} - \left[\left(\overline{\tau}_i^H + \overline{\tau}_i^E\right)\cdot\mathbf{n}_{ij}\right]_{\Pi_{ij}} = \gamma_{ij}\left(\overline{\nabla}\cdot\mathbf{n}_{ij}\right)\mathbf{n}_{ij} \quad \text{at } \overline{r} = \overline{r}_{\Pi_{ij}}(\theta,\overline{t}). \tag{7}$$

In equation (7), the electric and hydrodynamic stress tensors are denoted by $\overline{\tau}^E$ and $\overline{\tau}^H$ respectively.

### *2.3. Mathematical complexities and solution methodology*

First of all, we have chosen appropriate characteristic scales for representing governing equations as well as boundary conditions in a non-dimensional form. Those scales are: length ~ $\overline{a}_2$, velocity $(\overline{u}_c) \sim \varepsilon_3\overline{E}_\infty^2\overline{a}_2/\mu_3$, time ~ $\overline{a}_2/\overline{u}_c$, viscous stress ~ $\mu_3\overline{u}_c/\overline{a}_2$. The velocity scale is obtained by balancing the electric stress and hydrodynamic stress at the interface of the droplet. $\overline{E}_\infty$, and $\varepsilon_3\overline{E}_\infty^2$ have been used as characteristic scales for the applied electric field and electric stress respectively. After non-dimensionlization, we have obtained some important non-dimensional numbere. Those are: Reynolds number, $Re=\rho\varepsilon_3\overline{E}_\infty^2\overline{a}_2^2/\mu_3^2$ (relative strength of the inertia force force with respect to the viscous force acting on the fluid), electric capillary number, $Ca_E=\varepsilon_3\overline{E}_\infty^2\overline{a}_2/\gamma_{23}$ (which denotes relative strength of electric stress over the capillary stress); confinement ratio, $Wc=\overline{a}_2/\overline{H}$ (ratio of outer droplet diameter and total channel height) and radius ratio, $K=\overline{a}_1/\overline{a}_2$. Beside this, we have also identified some important property ratios: permittivity ratio, $S_{ij}=\varepsilon_i/\varepsilon_j$; conductivity ratio, $R_{ij}=\sigma_i/\sigma_j$ and viscosity ratio, $\lambda_{ij}=\mu_i/\mu_j$. One must acknowledge that the present mathematical problem is non-linear and coupled that restricts us to obtain its exact analytical solution for any arbitrary value of the non-dimensional parameters in an unbounded domain. Further complexity arises when the domain confinement is taken into consideration. Keeping this information in mind, we have taken the help of numerical simulation for examining the essential physics of the EHD of compound droplet in a highly confined channel. In addition, an asymptotic analysis is performed to support the numerical outcomes for the limiting scenario of low confinement and concentric compound droplet. The details of the analytical solution are presented in appendix A.

Towards performing a numerical simulation of the present problem, we have employed the diffuse interface based phase field method (Badalassi et al. 2003; Jacqmin 1999). One must acknowledge that, in phase field method, there is no necessasity of tracking the droplet surface as the sharp interface is replaced by the diffuse interface. In the present numerical model, we have considered that the inner droplet phase (core) and carrier fluid phase are same. In the phase field method, two immscible fluid phase in a binary system is represented by an order parameter $\phi$. For fluid 1 (inner droplet) and fluid 3 (suspending fluid), we have taken $\phi(\bar{\mathbf{x}}, \bar{t}) = -1$, whereas $\phi(\bar{\mathbf{x}}, \bar{t}) = 1$ is taken for fluid 2 (outer droplet). At the interface, the value of $\phi(\bar{\mathbf{x}}, \bar{t})$ varies from -1 to 1 rapidly. The order parameter $\phi(\bar{\mathbf{x}}, \bar{t})$ is regulated by the Cahn-Hilliard equation in the following form

$$\frac{\partial \phi}{\partial \bar{t}} + \bar{\mathbf{u}} \cdot \bar{\nabla} \phi = \bar{\nabla} \cdot (\bar{M}_\phi \bar{\nabla} \bar{G}), \tag{8}$$

where $\bar{M}_\phi$ and $\bar{G} = \gamma(\phi^3 - \phi)/\bar{\zeta} - \gamma \bar{\zeta} \bar{\nabla}^2 \phi$ represent the interface mobility factor and chemical potential factor respectively. Here, the parameter $\bar{\zeta}$ regulates the thickness of the interface. The Cahn-Hilliard equation in non-dimensional form reads as (Mandal et al. 2015; Bandopadhyay et al. 2016)

$$\frac{\partial \phi}{\partial t} + \mathbf{u} \cdot \nabla \phi = \frac{1}{Pe} \nabla^2 G. \tag{9}$$

The non-dimensional quantities in equation (9) are $G = (\phi^3 - \phi)/Cn - Cn \nabla^2 \phi$, which is the non-dimensional chemical potential, where $Cn = \bar{\zeta}/\bar{a}_2$ is known as Cahn number that controls the interface thickness. Another important non-dimensional parameter is the Péclet number denoted as $Pe = \bar{a}_2^2 \bar{u}_c / \bar{M}_\phi \gamma$. The distribution of the electric potential is obtained by solving the electric field equation depicted below (Reddy & Esmaeeli 2009)

$$\nabla \cdot (\sigma \nabla \varphi) = 0. \tag{10}$$

The pressure field and velocity field can be obtained by solving the continuity and Navier-Stokes equation, expressed as follows

$$\nabla \cdot \mathbf{u} = 0, \quad Re\left(\frac{\partial \mathbf{u}}{\partial t} + \nabla \cdot (\mathbf{uu})\right) = -\nabla p + \nabla \cdot \left[\mu\left\{\nabla \mathbf{u} + (\nabla \mathbf{u})^T\right\}\right] + G \nabla \phi + Ca_E \mathbf{F}^E. \tag{11}$$

In equation (11), $\mathbf{F}^E = \nabla \cdot (\varepsilon \nabla \varphi) \nabla \varphi - |\nabla \varphi|^2 \nabla \varepsilon / 2$ represents the volumetric electric force that is responsible for any interfacial deformation. It is worthy to note from equation (11) that the volumetric electric force is present in the governing Navier-Stokes equation in the diffuse interface framework unlike the sharp interface limit. The term $G \nabla \phi$ represents the phase field dependent interfacial tension force. Under the phase field framework, the fluid properties are represented in the following form

$$\left. \begin{array}{l} \rho = \dfrac{(1+\phi)}{2} \rho_r + \dfrac{(1-\phi)}{2}, \mu = \dfrac{(1+\phi)}{2} \lambda + \dfrac{(1-\phi)}{2} \\ \varepsilon = \dfrac{(1+\phi)}{2} S + \dfrac{(1-\phi)}{2}, \sigma = \dfrac{(1+\phi)}{2} R + \dfrac{(1-\phi)}{2}. \end{array} \right\} \quad (12)$$

For numerical simulation using phase field method, we have employed the finite element based COMSOL Multiphysics software.

## 3. Results and discussions

### 3.1. Comparison between numerical and analytical results

First of all, we have made a comparison between our theoretical prediction and the results obtained from the full scale numerical simulation. We have performed the theoretical analysis in an unbounded domain, the details of which is shown in appendix A. In order to attain a greater accuracy in our analytical prediction, we have considered higher order correction terms in droplet shape. Based on small-deformation perturbation analysis, the inner and outer droplet shape can be represented as (Mandal, Ghosh, et al. 2016)

$$\left. \begin{array}{l} r_{s_{12}}(\theta) = K\{1 + f_{12}(\theta)\} = K\left\{1 + Ca_E f_{12}^{(Ca_E)} + Ca_E^2 f_{12}^{(Ca_E^2)} + O(Ca_E^3)\right\}, \\ r_{s_{23}}(\theta) = \{1 + f_{23}(\theta)\} = \left\{1 + Ca_E f_{23}^{(Ca_E)} + Ca_E^2 f_{23}^{(Ca_E^2)} + O(Ca_E^3)\right\}. \end{array} \right\} \quad (13)$$

Here $f_{ij}^{(Ca_E)}$ and $f_{ij}^{(Ca_E^2)}$ denote the corrections in shape of the interface *ij* at different orders of perturbation, namely $O(Ca_E)$ and $O(Ca_E^2)$ respectively. The corrections can be expressed in the following form as

$$O(Ca_E): \quad f_{12}^{(Ca_E)} = M_{2,0}^{(Ca_E)} \cos(2\theta), \quad f_{23}^{(Ca_E)} = L_{2,0}^{(Ca_E)} \cos(2\theta), \quad (14)$$

$$O(Ca_E^2): \begin{cases} f_{12}^{(Ca_E^2)} = M_{0,1}^{(Ca_E^2)} + \left[ M_{2,1}^{(Ca_E^2)} \cos(2\theta) + M_{4,1}^{(Ca_E^2)} \cos(4\theta) \right], \\ f_{23}^{(Ca_E^2)} = L_{0,1}^{(Ca_E^2)} + \left[ L_{2,1}^{(Ca_E^2)} \cos(2\theta) + L_{4,1}^{(Ca_E^2)} \cos(4\theta) \right], \end{cases} \quad (15)$$

where $L_{0,1}^{(Ca_E^2)}$ and $M_{0,1}^{(Ca_E^2)}$ can again be expressed as

$$L_{0,1}^{(Ca_E^2)} = -\frac{1}{4}\left\{L_{2,0}^{(Ca_E)}\right\}^2 - \frac{1}{4}\left\{\hat{L}_{2,0}^{(Ca_E)}\right\}^2, \quad M_{0,1}^{(Ca_E^2)} = -\frac{1}{4}\left\{M_{2,0}^{(Ca_E)}\right\}^2 - \frac{1}{4}\left\{\hat{M}_{2,0}^{(Ca_E)}\right\}^2. \quad (16)$$

The constant coefficients in the above equations are provided in appendix A. The deformation of both of the interfaces can be quantified in terms of a deformation parameter, which can be expressed in the following form

$$D_{ij} = \frac{r_{\Pi_{ij}}(\theta = \pi/2) - r_{\Pi_{ij}}(\theta = 0)}{r_{\Pi_{ij}}(\theta = \pi/2) + r_{\Pi_{ij}}(\theta = 0)}. \quad (17)$$

In the present analysis, we have considered two PD-PD-PD (all three phases are perfect dielectric in nature) systems with $(S_{23},S_{12})=(10,1)$ and $(S_{23},S_{12})=(0.1,10)$. Three LD-LD-LD systems (here all three phases are leaky dielectric in nature) are also considered where system I is characterized by $(S_{23}, R_{23})=(0.4397, 0.033)$ & $(S_{12}, R_{12})=(2.274, 30.33)$, system II has $(S_{23}, R_{23})=(1, 10)$ & $(S_{12}, R_{12})=(1, 0.1)$ and finally system III has $(S_{23}, R_{23})=(2, 0.5)$ & $(S_{12}, R_{12})=(0.5, 2)$. These parameters values of $R_{ij}$ and $S_{ij}$ have been extracted from previously performed experimental and numerical studies (Torza et al ., 1970; Tsukada et al ., 1994; Mählmann and Papageorgiou, 2009). It is important to mention that the analytical prediction for the steady-state deformation parameter ($D_\infty$) is a obtained by applying the limit $t\to\infty$. On the other hand, the numerical simulations are run for a prolong time ($t\sim 50$) to achieve $D_\infty$. In order to obtain a good match with our theoretical prediction, we have chosen $Wc=0.2$, which nearly approximates an unbounded flow field. It is made sure that the numerical results do not depend on grid size and Cahn number (details the same is provided in Appendix B).

In figure 2, we have shown the variation of steady-state deformation parameter ($D_\infty$) with $Ca_E$ both numerically and analytically. Figure 2(a) and 2(b) depict the variation of $D_{\infty,23}$ and $D_{\infty,12}$ with $Ca_E$ for the outer and inner droplet of a PD-PD-PD model whereas figure 2(c) & 2(d) show the corresponding variation for a LD-LD-LD model (system I). Droplet shapes corresponding to different values of $Ca_E$ are also shown in figure 2. These plots in figure 2 illustrate that our higher order [$O(Ca_E^2)$ correction in shape] asymptotic theory exhibits a much better match with the numerically obtained results as compared to the leading order theoretical prediction.

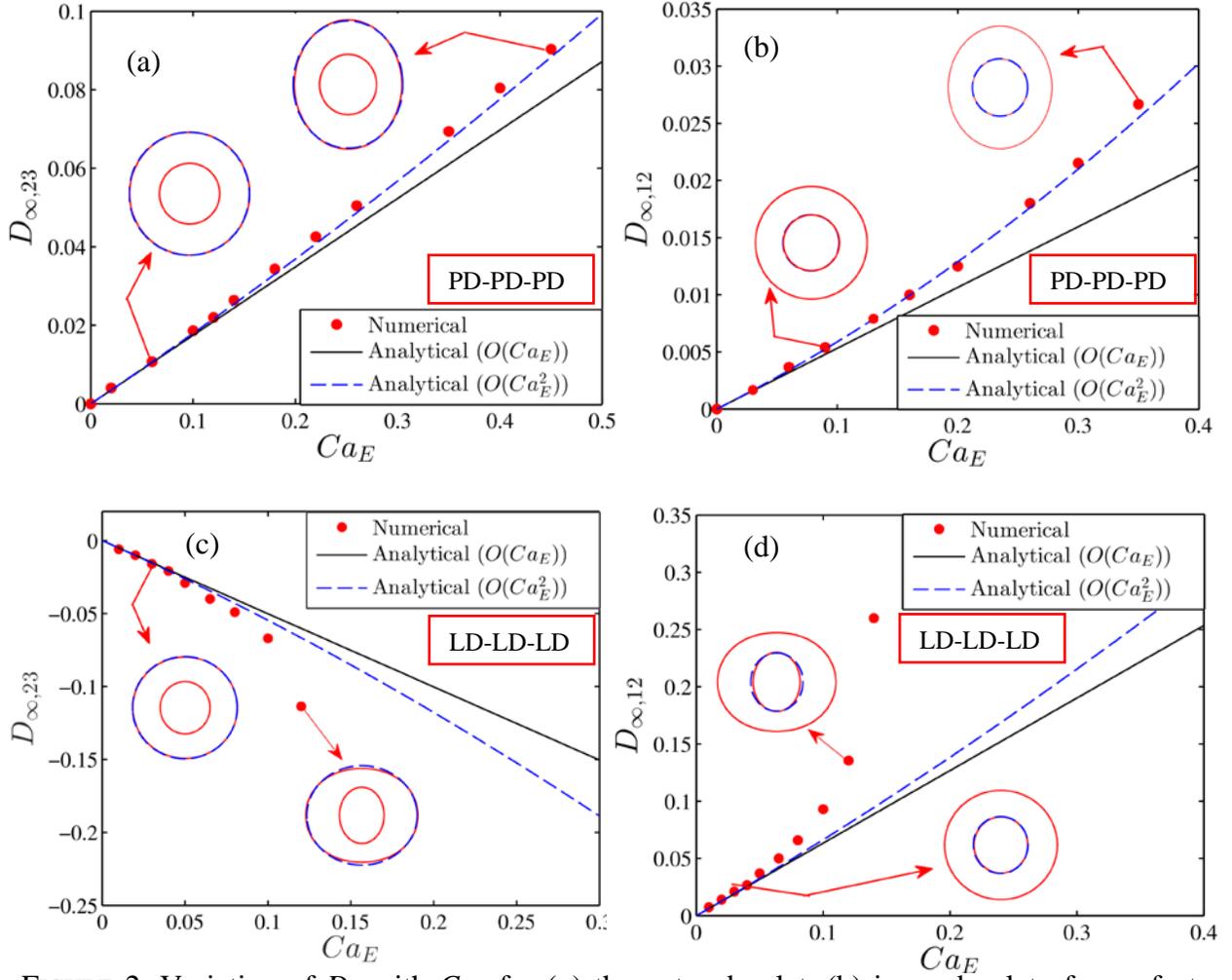

FIGURE 2. Variation of $D_\infty$ with $Ca_E$ for (*a*) the outer droplet, (*b*) inner droplet of a perfect dielectric system with $(S_{12}, S_{23})$=(10, 0.1) (*c*) outer droplet and (*d*) inner droplet of system I with $(S_{23}, R_{23})$=(0.4397, 0.033) & $(S_{12}, R_{12})$=(2.274, 30.33). Other used parameters are $Re = 0.01$, $\lambda=1$ and $e=0$.

### 3.2 Channel confinement-induced variation in steady state deformation

Figure 3 shows the effect of channel confinement on the steady state deformation of a compound droplet for a PD-PD-PD system with $(S_{23}, S_{12})$ =(10, 0.1). The channel confinement ratio is varied between $Wc = 0.2$ to $Wc = 0.8$ to obtain a better understanding of the effect of bounding wall. From figure 3(a), it can be observed that the steady state deformation of both the inner and outer droplet increases with rise in the channel confinement. Both the interfaces are deformed into a prolate shape, which is evident from figure 3(b). The primary reason for such a behavior is due to the rise in strength of the electric field with increase in the domain confinement. Figure 3(c) shows that the magnitude of electric field strength is high for a highly confined domain. Therefore, the magnitude of normal electric stress is also high which results in an enhanced deformation of the droplet. It is also important to note that the outer droplet undergoes a greater droplet deformation as compared to the inner droplet. Such a behavior can be attributed to the following two reasons (i) the permittivity of the inner droplet is low and as a result lower Maxwell stress is generated at the interface, (ii) the restoring force of the inner droplet is much more significant due to lower interfacial curvature. Another important aspect to be noted is that the deformation of the outer droplet is less in comparison to a single droplet. As the permittivity of the inner droplet is less, the Maxwell stress generated at the inner interface is small that creates a lower deformation of the interface. Therefore, the intensity of the viscous flow (and hence the viscous stress) at the annular region, generated due to the interface deformation is also weak, which causes lower deformation of both the interfaces.

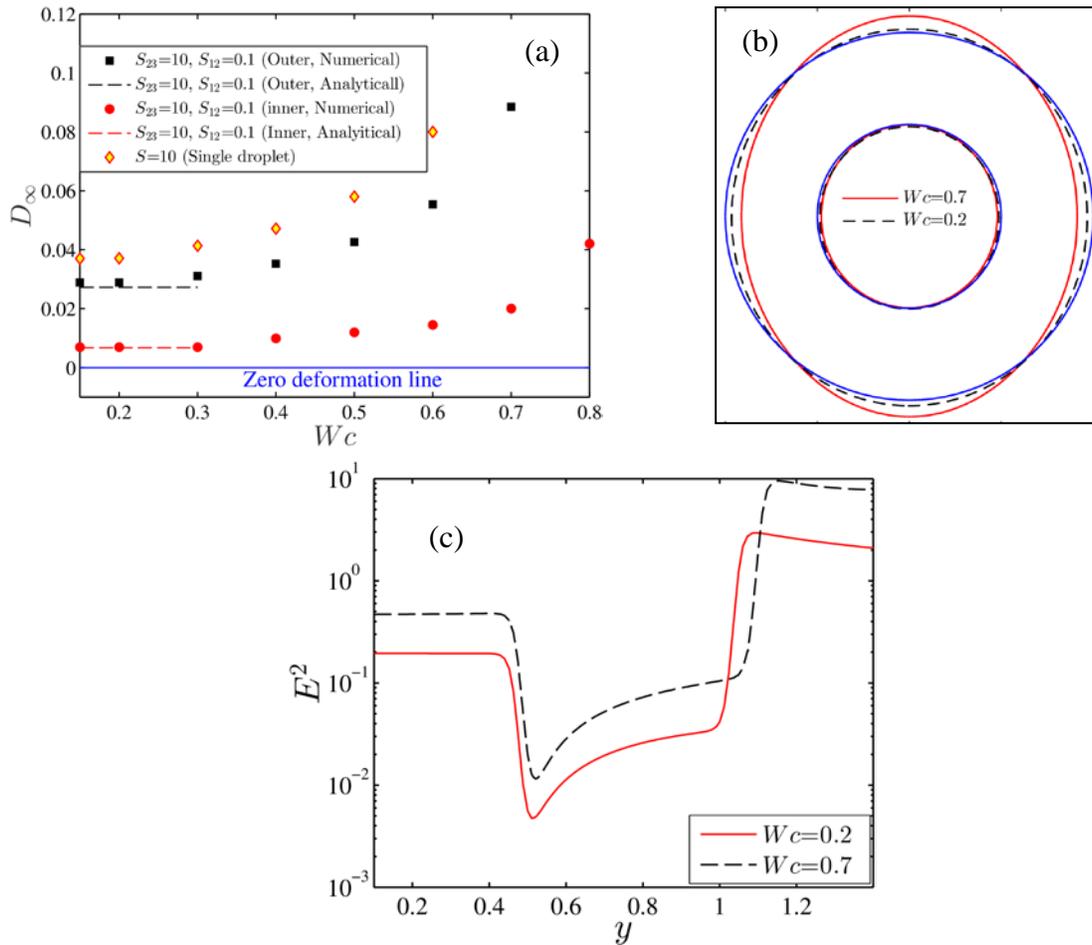

FIGURE 3. (a) Variation of $D_\infty$ with $Wc$ for a PD-PD-PD system at $Ca_E$ =0.15, (b) droplet shapes for different values of $Wc$ where the blue colored contour shows the undeformed droplet shape, (c) variation of $E^2$ with $Wc$ along a vertical line drawn from the droplet center to the wall electrode. Other parameters are $(S_{23}, S_{12})$ =(10, 0.1), $Re$=0.01, $\lambda$ =1, $e$=0.

Similarly, figure 4 demonstrates the variation of steady state deformation parameter as a function of the domain confinement for a leaky dielectric system II with $(R_{23}, S_{23})$=(10, 1) and $(R_{12}, S_{12})$=(0.1, 1). The figure demonstrates that the steady state deformation of the inner droplet reduces with rise in the confinement ratio, whereas the deformation of the outer droplet increases till $Wc$=0.80. Any further increase in the channel confinement results in a reduction in the deformation of the outer droplet. For a leaky dielectric system, the deformation dynamics of the droplet is mainly regulated by the interplay between normal electric and hydrodynamic stress. In an unbounded domain, these stresses act in the same direction and try to deform the outer droplet ($R_{23}$>$S_{23}$) and inner droplet ($R_{12}$<$S_{12}$) into prolate and oblate configuration respectively. It is worthy to mention that the confinement effect induces higher electric potential in a confined domain and increases the electric field strength similar to a PD-PD-PD system as depicted in figure 4(b). Therefore, with the rise in the domain confinement, the normal electric stress increases that enhances the deformation.

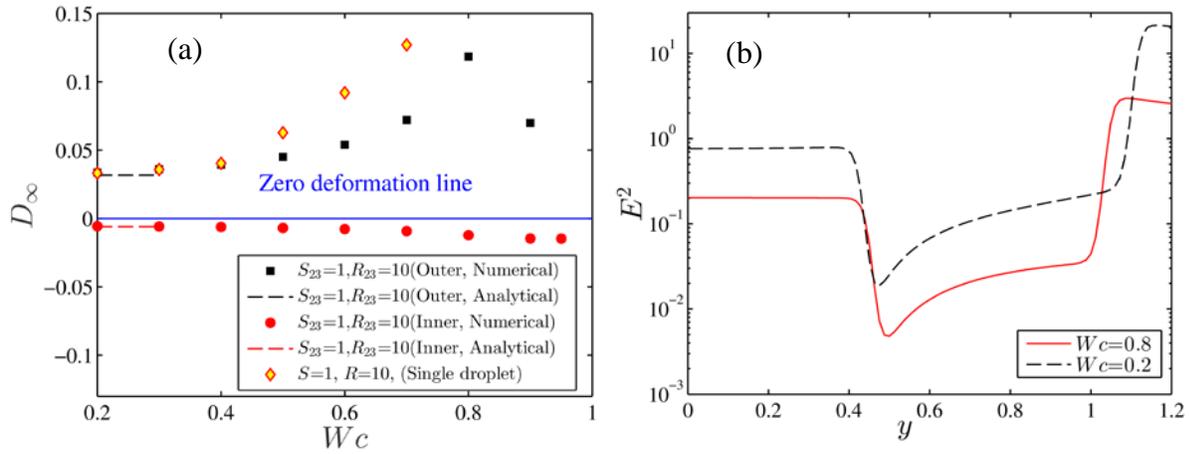

FIGURE 4 (a) Variation of $D_\infty$ with $Wc$ for system II with ($S_{23}$, $R_{23}$)=(1, 10) and ($S_{12}$, $R_{12}$)=(1, 0.1) at $Ca_E$ =0.1, (b) variation of $E^2$ with $Wc$ along a vertical line drawn from the droplet center to the wall electrode. Other used parameters are $Re$ = 0.01, $\lambda$=1 and $e$=0.

On the contrary, the magnitude of the normal hydrodynamic stress decreases and it reverses when the confinement ratio increases beyond a certain critical value.. We have described it as critical confinement ratio ($Wc_{critical}$). Further increment of the channel confinement increases the magnitude of reversed hydrodynamic stress that now tries to oppose the effect of the normal electric stress. For the outer droplet, as the value of $R_{23}$ is high, the effect of domain confinement on the normal electric stress is also high and it dominates the decrease in normal hydrodynamic stress till $Wc$=0.80. Increase in $Wc$ beyond the value of 0.8 reverses the direction of the normal hydrodynamic stress as well as increases its magnitude significantly such that a reduction in shape deformation of the outer interface results. For the present LD-LD-LD model too, the deformation of the inner droplet is lower as compared to the outer droplet because of the lower conductivity and a higher restoring force of the inner droplet. Like the PD-PD-PD model, the steady state deformation of the outer droplet of the present system is smaller than that of a single droplet due to the presence of the inner droplet.

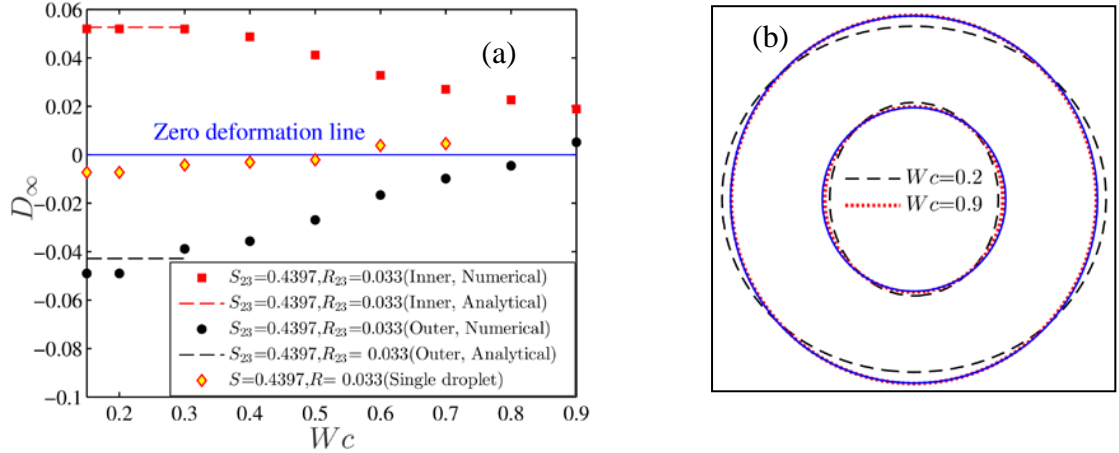

FIGURE 5. (a) Variation of $D_\infty$ with $Wc$ for system I with $(S_{23}, R_{23})=(0.4397, 0.033)$ and $(S_{12}, R_{12})=(2.274, 30.33)$ at $Ca_E = 0.08$, (b) droplet shapes for different values of $Wc$ where the blue colored contour shows the undeformed droplet shape. Other used parameters are $Re = 0.01$, $\lambda=1$, and $e=0$.

Similarly, figure 5 shows the alteration of steady state deformation parameter with domain confinement for system I having $(R_{23}, S_{23})=(0.033, 0.4397)$ and $(R_{12}, S_{12})=(30.33, 2.274)$. From figure 5(a), it can be noted that the inner and outer droplet deforms into a prolate and a oblate configuration, respectively, in an unbounded domain. With the rise in channel confinement, the steady state value of the deformation parameter of both the interfaces (inner and outer) decreases. When the droplet is highly confined in the channel that is $Wc \geq 0.8$, both the interfaces deform into a prolate configuration. This phenomenon happens because of the critical interplay between normal electric and normal hydrodynamic stresses in a confined domain. In case of a leaky dielectric system, the strength of normal electric stress increases with domain confinement similar to the system II. On the other hand, the normal hydrodynamic stress decreases and above a critical domain confinement, it gets reversed, as mentioned in the study of Esmaeeli and Behjatian (2013a) & Santra et al. (2018). For the outer droplet, the increase in normal electric stress with domain confinement is less as compared to the reduction in normal hydrodynamic stress as the conductivity ratio is low ($R_{23}$). So, the steady state deformation of the outer droplet reduces and beyond $Wc \approx 0.80$, the reversed-normal hydrodynamic stress dominates the normal electric stress that leads the droplet to deform into prolate configuration. It should also be noted that the critical confinement ratio ($Wc_{critical} \approx 0.8$) is higher as compared to a single droplet ($Wc \approx 0.5$) due to the presence of inner droplet. For the present LD-LD-LD model, the surface charge induced at the inner droplet interface is higher due to the large magnitude of $R_{12}$. So, the viscous flow generated due to the disparity in the tangential electric stresses is also high that increases the deformation of both the interfaces. In addition, the Maxwell stress induced at the inner droplet interface is also large that creates a significantly higher prolate deformation of the inner droplet. As a consequence, the flow in the annular region gets squeezed which increases the strength of the viscous stress at both the interfaces and further enhances the deformation. Hence, the oblate deformation of the outer droplet is significantly higher as compared to a single droplet and the transformation of droplet shape from oblate to prolate configuration takes place at comparatively higher domain confinement ratios.

*3.3 Channel confinement-induced variation in transient deformation characteristics*

Figure 6 depicts the temporal evolution of the deformation parameter for different values of $Ca_E$ in a weakly confined domain ($Wc=0.2$). One must acknowledge that time is normalized with the scale $\bar{a}_2/\bar{u}_c$ in all the transient plots. In these plots, we have plotted the numerical data and compared it with the theoretical predictions.

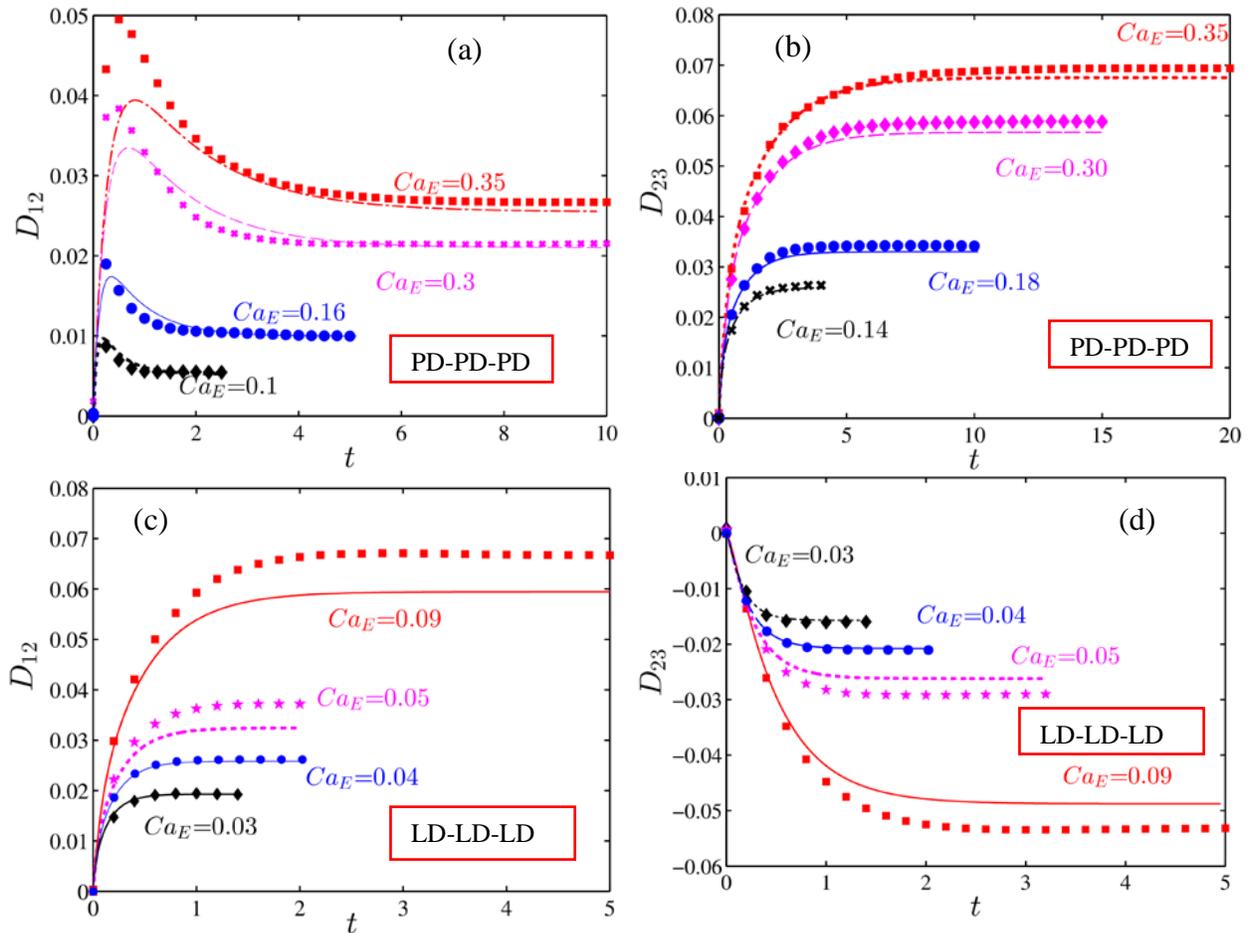

FIGURE 6. Temporal variation of deformation parameter for (*a*) the inner droplet, (b) outer droplet of a PD-PD-PD system with ($S_{12}$, $S_{23}$)=(10, 0.1) (*c*) inner droplet, (*d*) outer droplet of system I with ($S_{23}$, $R_{23}$)=(0.4397, 0.033) and ($S_{12}$, $R_{12}$)=(2.274, 30.33). Other parameters are $Wc=0.2$, $Re = 0.01$, $\lambda=1$ and $e =0$. In these plots, the markers show the numerical results and the lines show the analytical results

Figure 6(a) and 6(b) show the variation of $D_{12}$ as a function of time for the inner and outer droplets of a PD-PD-PD system with ($S_{12}$, $S_{23}$)=(10, 0.1), whereas figures 6(c) and 6(d) show the variation of the corresponding deformation in a LD-LD-LD system (system I). At lower values of $Ca_E$, all the plots show a good match between the analytical and the numerical results. However, at higher values of $Ca_E$, the analytical solution under-predicts the deformation for both the inner and outer droplet for a PD-PD-PD system. On the other hand, the analytical solution over-predicts the deformation for the outer droplet whereas under-predicts it for the inner droplet for the system I.

Further, we have also shown the effect of domain confinement on the time required to attain the steady state shape (or steady state time) of the inner and outer droplet of a PD-PD-PD with $(S_{23}, S_{12}) = (10, 0.1)$ and LD-LD-LD system (system II) in figure 7.

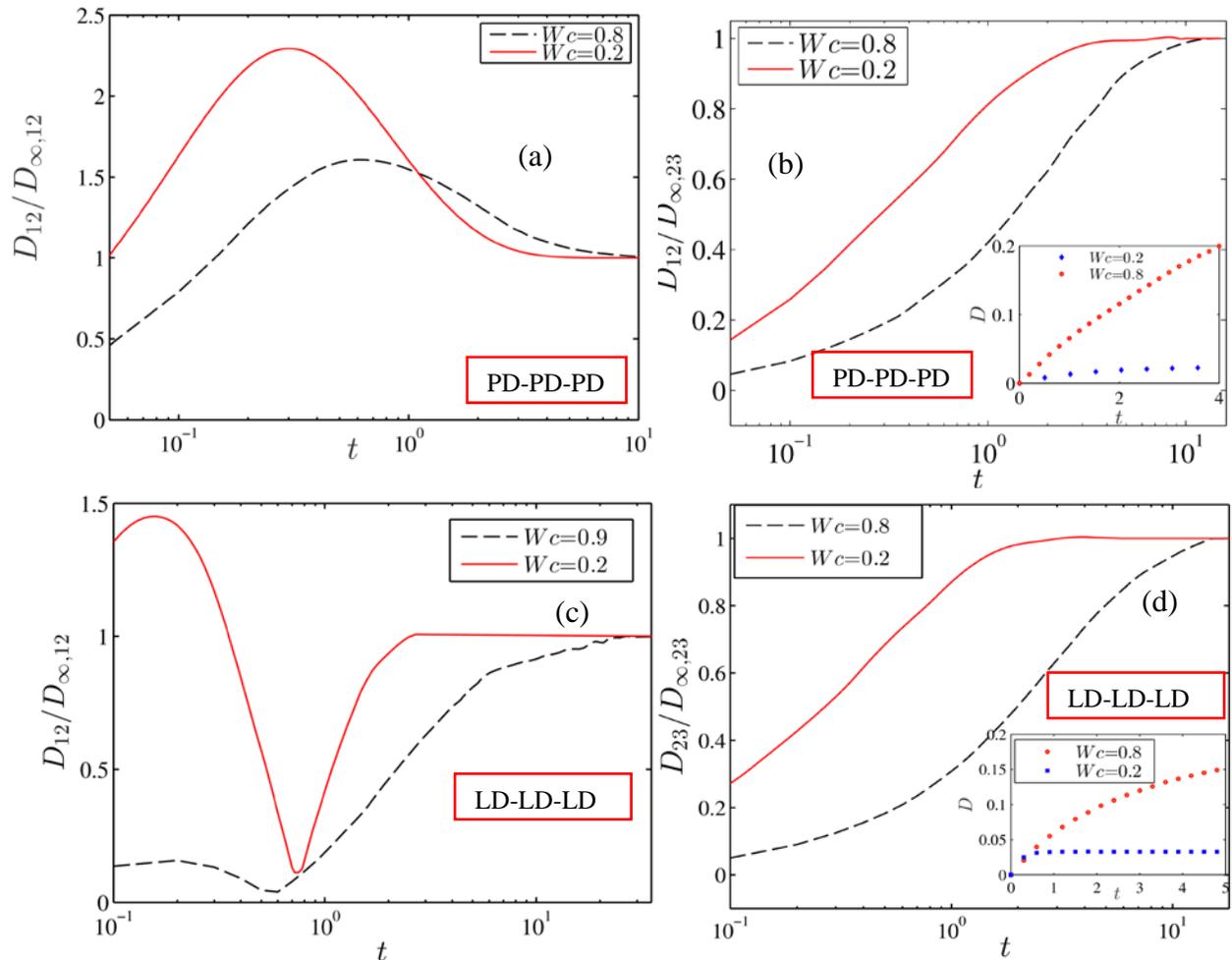

FIGURE 7. Impact of confinement on the steady state time for (a) the inner droplet, (b) outer droplet of a PD-PD-PD system with $(S_{23}, S_{12}) = (10, 0.1)$ and (c) inner droplet, (d) outer droplet of system II with $(S_{23}, R_{23}) = (1, 10)$ and $(S_{12}, R_{12}) = (1, 0.1)$. The insets of figure 7(b) and 7(d) show the variation of same for a single droplet. Other parameters are $Wc=0.2$, $Re = 0.01$, $e=0$ and $\lambda=1$.

Figure 7(a) & 7(c) show a non-monotonic variation of the deformation parameter of the inner droplet as a function of time elapsed for both a perfect and a leaky dielectric system,

respectively. In either of these systems, a weakly confined compound droplet reaches it steady state configuration earlier as compared to a tightly confined one. On the other hand, figure 7(b) & 7(d) show that the transient variation of the deformation parameter of the outer droplets of a PD-PD-PD and LD-LD-LD system. For both these systems, the deformation parameter undergoes a monotonic variation and the droplets achieve a steady state configuration very sluggishly in the presence of confinement. Important point to be noted is that a single droplet undergoes a drastic elongation that leads to its breakup for the same values of $R$, $S$ and $Ca_E$ unlike the outer interface of the compound droplet (refer to the inset of figure 7(b) and 7(d)). This behavior is primarily due to the lower conductivity (for a LD-LD-LD system) and permittivity (for a PD-PD-PD system) of the inner droplet. For the present LD-LD-LD system, as the value of $R_{12}$ is low, the Maxwell stress generated at the interface is also less that results in lower deformation. Furthermore, the strength of the viscous flow generated due to the jump in the tangential electrical stress is also low which reduces the deformation of the interface. Due to the same reason, the outer droplet is able to sustain its steady state configuration at comparatively higher values of $Ca_E$, unlike a single droplet. A similar explanation is also applicable for the perfect dielectric system, where the presence of the inner droplet reduces the deformation of the outer droplet and helps the outer droplet to achieve steady state configuration at comparatively higher value of $Ca_E$.

### *3.4. Effect of eccentricity of the inner droplet on the its motion*

This section is devoted to investigate the motion of an eccentrically placed inner droplet in the presence of a uniform electric field. Figure 8(a) shows the impact of domain confinement on the movement of the inner droplet of system II with $(S_{23},R_{23}) = (1,10)$ and $(S_{12},R_{12}) = (1,0.1)$. From the figure, it can be observed that the inner droplet moves at a faster rate in a confined domain as compared to an unbounded domain. This translational motion of the inner droplet depends on imbalance of electrohydrodynamic force (EHD force) created due to the asymmetric distribution of fluid flow around the inner droplet. The net EHD force can shift the inner droplet towards the east pole ($\theta=0$) or the west pole ($\theta=\pi$) of the outer droplet based on the direction of fluid flow whose magnitude again depends on the strength of the flow circulation. The direction of fluid flow is from the poles to equators of the inner droplet as shown in figure 8(b), for the values of $R$ and $S$ considered ($S_{12}>R_{12}$). So, the electrohydrodynamic force at the east pole of the inner droplet ($\theta = 0$) tries to pull the droplet towards right whereas the force at the west pole ($\theta = \pi$) forces it to move towards left. Now, figure 8(c) shows that the peak value of magnitude of velocity ($U$) of the inner droplet is higher at the left of the inner droplet, which necessarily signifies a greater strength of the fluid flow circulation in the same region. As a result, the net EHD force drives the inner droplet towards the left. In a confined domain, the deformation is more and the strength of fluid flow circulation is also high, as depicted in figure 8(c), which creates a net EHD force having greater strength, that in turn causes a faster migration of the inner droplet.

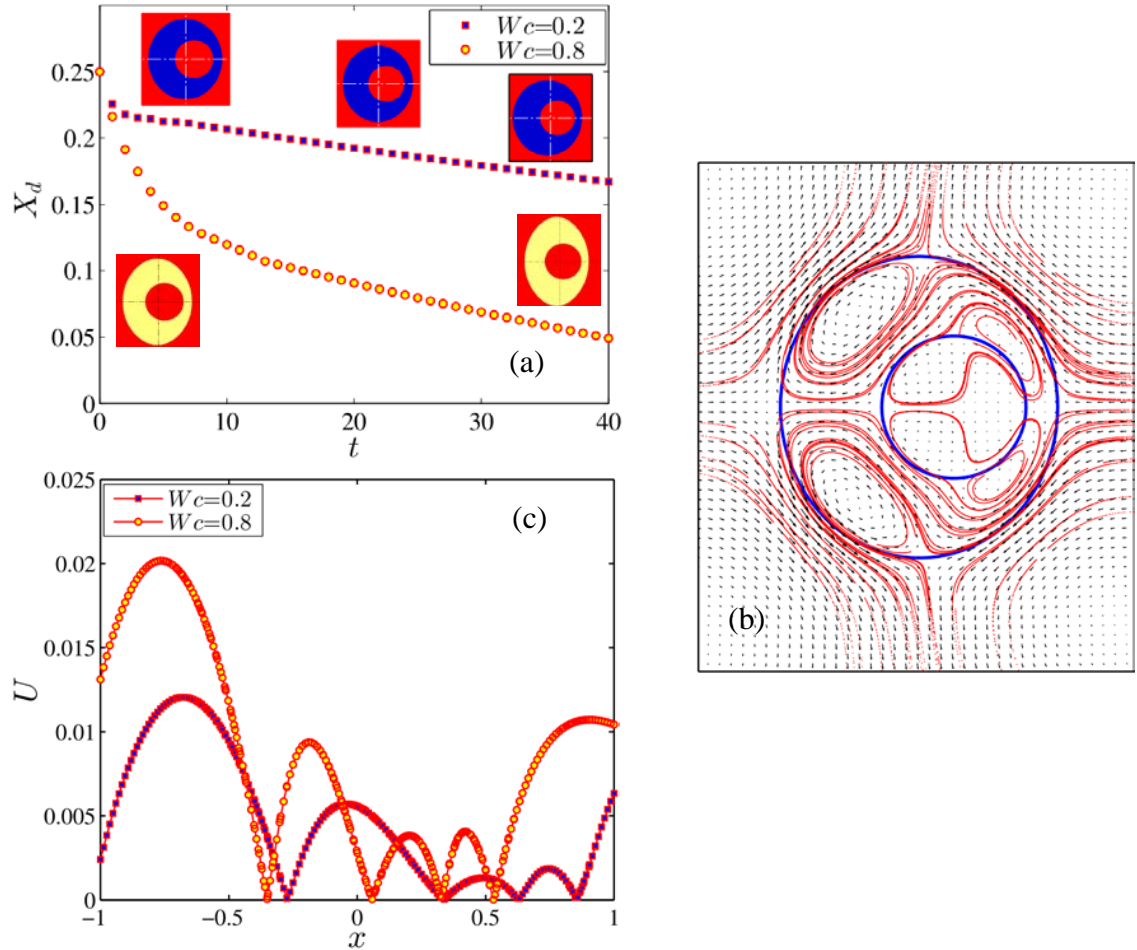

FIGURE 8. (a) Impact of confinement on the temporal variation of the position of the inner droplet for system II with $(S_{23},R_{23}) = (1,10)$, $(S_{12},R_{12}) = (1,0.1)$, (b) streamline pattern of fluid flow circulation at $Wc=0.2$, (c) effect of confinement on variation of velocity along straight horizontal line passing through the center of the droplets from west pole to east pole of the outer droplet at $t=10$. Other parameters are $Re = 0.01$, $Ca_E=0.125$, $e=0.25$ and $\lambda=1$.

Similarly, figure 9(a) shows the impact of channel confinement on the translational motion of the inner droplet of system III with $(S_{23},R_{23}) = (2,0.5)$ and $(S_{12},R_{12}) = (0.5,2)$. Figure 9(a) displays the fact that the inner droplet undergoes a to and fro motion for a weakly confined droplet. However, this to and fro motion surprisingly vanishes in a highly confined domain and the inner droplet moves towards west pole of the outer droplet at comparatively lower speed. For the values of $R$ and $S$ ($S_{12}<R_{12}$) used, the inner droplet undergoes a prolate deformation and the fluid flow occurs from equator to pole of the inner droplet as shown in figure 9(b). Therefore, the electrohydrodynaimic force at the west pole tries to push the droplet towards the east pole. On the contrary, the EHD force at the east pole tries to push the inner droplet in the opposite direction. Initially ($t = 1$), the strength of fluid flow circulation is high at the left of the inner droplet as can be seen from figure 9(c). Therefore, the net EHD force initially drives the droplet towards the east pole of the outer droplet. Due to the movement of inner droplet towards right of it, the gap between the inner droplet and outer droplet reduces. This results in an increase in the film pressure at the reduced gap which further creates a repulsive force that forces the inner droplet towards west pole of the outer droplet. When the gap between the inner and outer droplet increases sufficiently, the net EHD force again dominates and pushes the inner droplet towards the east pole of the outer droplet. This interplay between the repulsive force and net EHD force creates a to and fro motion of the inner droplet. Figure 9(d) shows that the strength of fluid flow circulation is less in confined domain resulting in a reduced strength of the net EHD force. Therefore, the translational motion of the inner droplet is dominated by the film pressure due the presence of the outer droplet.

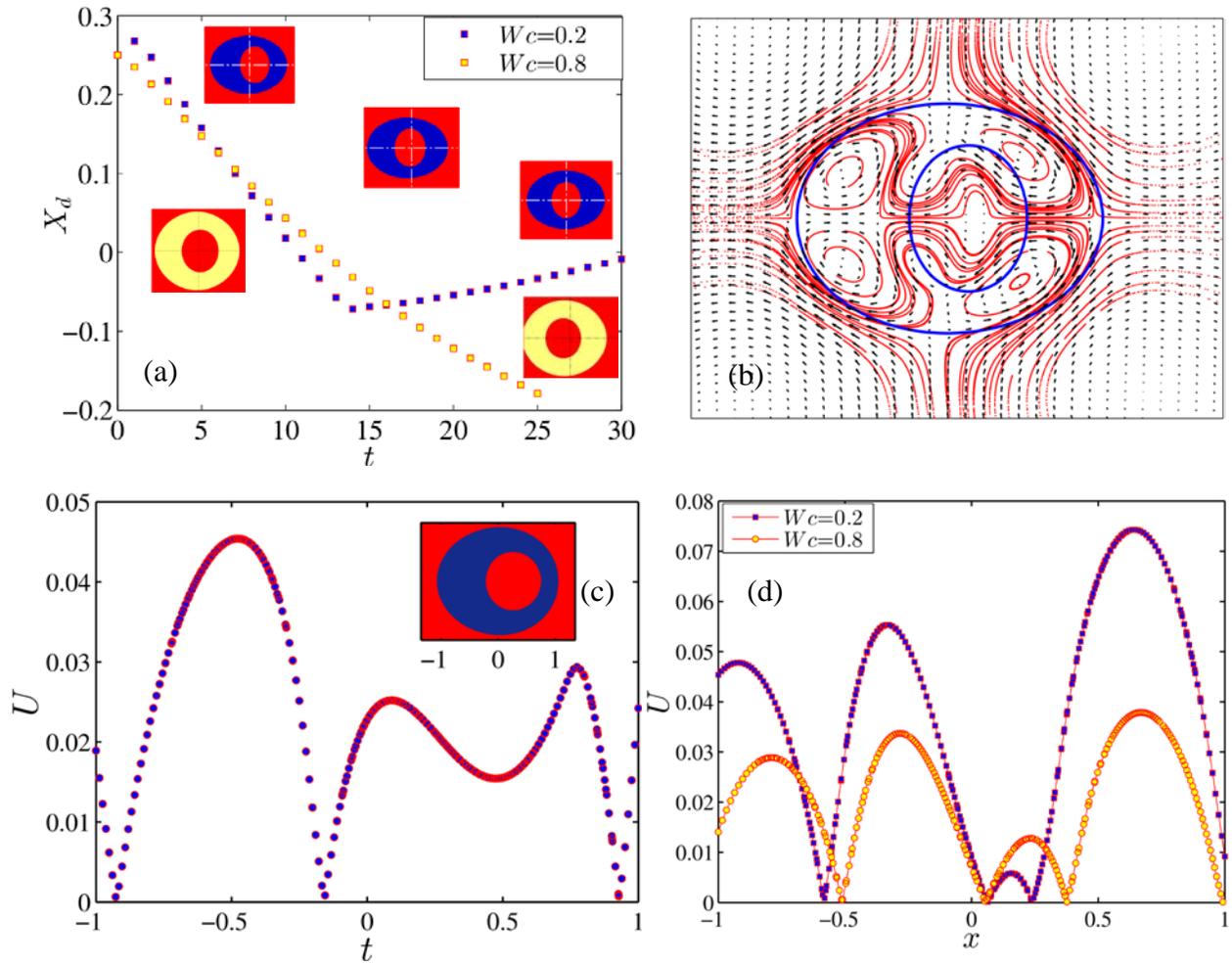

FIGURE 9.(a) Effect of confinement on the temporal variation of the position of the inner droplet for system III with $(S_{23}, R_{23}) = (2, 0.5)$, $(S_{12}, R_{12}) = (0.5, 2)$, (b) streamline pattern of fluid flow circulation in and around the droplet at $Wc = 0.2$, (c) variation of magnitude of velocity along a straight line passing through the cente from west pole to east pole of the droplet at $t=1$, (d) effect of confinement on the magnitude of velocity along a horizontal straight line passing through the center of the droplet from east pole to west pole. Other parameters are $Ca_E=0.20$, $Re = 0.01$, $e = 0.25$, and $\lambda=1$.

Next, we have formed a regime diagram that shows the pattern of the droplet motion based on the values of ($R$, $Wc$) in figure 10. We have obtained two different patterns of inner droplet motion. The yellow colored region with diamond-shaped markers indicates that the droplet undergoes to and fro motion, whereas the blue-colored regime with triangular-shaped marker signifies a translational motion of the inner droplet towards the center of the outer droplet.

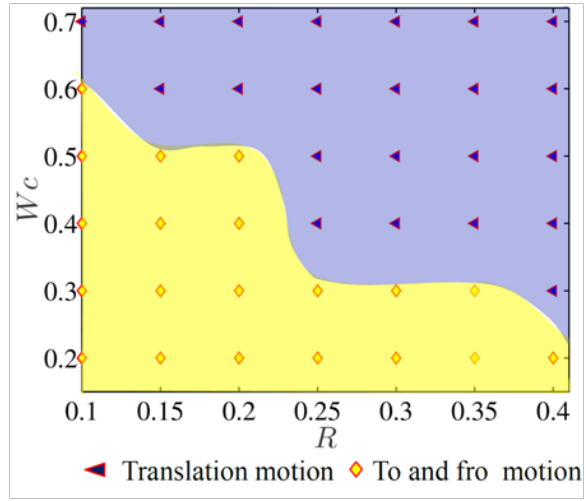

FIGURE 10. Regime plot showing the motion characteristics of the inner droplet. The different parameters are ($S_{23}$,$S_{12}$) = (0.5,2). $\lambda = 1$, $Ca_E = 0.2$ and $e = 0.1$. In the present plot, we have considered that $R_{23}=R$.

## 4. Conclusions

In the present study, we have examined the impact of channel confinement on the electrohydrodynamics of a compound droplet in presence of a uniform DC electric field. We have performed 2D numerical simulations to explore the underlying physics of droplet dynamics in a confined domain. We have also performed a higher-order asymptotic analysis in order to validate the simulation results for the limiting case of low channel confinement. Throughout this study, we have shed light on the impact of channel confinement on both the transient and steady state deformation characteristics of a compound droplet as well as the motion of an eccentrically placed inner droplet. Some of the important results, worth highlighting, are presented below

1. For a PD-PD-PD system having $S_{23}>1$ and $S_{12}<1$, the steady state deformation of both the inner and outer droplet increase with the confinement ratio. However, the deformation is low for a compound droplet as compared to a single droplet.

2. For a LD-LD-LD system, the deformation of the inner droplet always shows a monotonic dependency on the confinement ratio. On the contrary, the outer droplet not only shows a non-monotonic variation in deformation, but also shows that the necessary condition for prolate deformation of the outer droplet changes from $R > S$ to $S > R$, if the confinement ratio increases beyond $Wc_{critical}$, the magnitude of which is more for a compound droplet as compared to a single droplet.

3. In a confined domain, both the inner and the outer droplet attains steady state configuration sluggishly for both the PD-PD-PD and LD-LD-LD system. The outer droplets of both these systems achieve steady state configuration in a confined domain, whereas a single droplet undergoes continuous elongation that leads to its breakup under the same circumstances.

4. In an unbounded domain, an eccentrically placed inner droplet of a LD-LD-LD system follows translational or to-and-fro migration characteristics depending on the value of $R$ and $S$. On the other hand, the dependency of this migration pattern on $R$ and $S$ ceases in a confined domain. At the same time, the translational migration of the inner droplet occurs at a faster rate in a confined channel.

**Appendix A. Asymptotic analysis for a concentric compound droplet shape in an unbounded domain**

In the creeping flow regime ($Re \ll 1$), the analytical solution of the current EHD problem in an unbounded domain can be achieved by regular perturbation method considering $Ca_E$ as the perturbation variable. Using perturbation method, we can get an analytical solution of the electrical potential and flow field. According to small deformation theory, the expansion of any field variable $\chi$ takes the following form (Vlahovska. P. M. 2011; Mandal & Chakraborty 2017)

$$\chi = \chi^{(0)} + Ca_E \chi^{(Ca_E)} + Ca_E^2 \chi^{(Ca_E^2)} + O(Ca_E^3). \tag{A1}$$

Here $\chi^{(0)}$ stands for the leading-order term of $\chi$, when shape deformation is not present and $\chi^{(Ca_E)}$ and $\chi^{(Ca_E^2)}$ symbolize the first and second order corrections to the spherical shape of the droplet. In this study, we have obtained a higher-order analytical solution (droplet shape correct upto $O(Ca^2{}_E)$) and have performed a comaprison with the numerical result.

The electric potential in the inner droplet, outer droplet and suspending medium satisfies Laplace equation and its general solution can be expressed as

$$\left. \begin{aligned} \varphi_1 &= \sum_{n=1}^{\infty} r^n \sum_{m=1}^{n} \left[ a_{n,m} \cos(m\theta) + \hat{a}_{n,m} \sin(m\theta) \right], \\ \varphi_2 &= \sum_{n=1}^{\infty} r^n \sum_{m=1}^{n} \left[ a_{n,m} \cos(m\theta) + \hat{a}_{n,m} \sin(m\theta) \right] + \sum_{n=1}^{\infty} \frac{1}{r^n} \sum_{m=1}^{n} \left[ e_{n,m} \cos(m\phi) + \hat{e}_{n,m} \sin(m\phi) \right], \\ \varphi_3 &= -\mathbf{E}_\infty \cdot \mathbf{r} + \sum_{n=1}^{\infty} \frac{1}{r^n} \sum_{m=1}^{n} \left[ b_{n,m} \cos(m\phi) + \hat{b}_{n,m} \sin(m\phi) \right], \end{aligned} \right\} \tag{A2}$$

where the unknown coefficients at each order of perturbation are obtained by applying the boundary conditions as follows

(i) electric potential is continuous at the fluid-fluid interface

(ii) normal component of current density is also continuous at the fluid-fluid interface.

In the Stokes flow limit, the stokes equation for $i$th fluid in terms of stream function can be written as

$$\nabla^4 \psi_i = 0 \qquad (A3)$$

The general solutions of the stream function inside and outside of the droplet can be written as

$$\begin{aligned}
\psi_1 &= \sum_{n=2}^{\infty} r^n \sum_{m=2}^{n} \left( A_{n,m} \cos(m\theta) + \hat{A}_{n,m} \sin(m\theta) + B_{n,m} r^2 \cos(m\theta) + \hat{B}_{n,m} r^2 \sin(m\theta) \right), \\
\psi_2 &= \sum_{n=2}^{\infty} r^n \sum_{m=2}^{n} \begin{pmatrix} F_{n,m} \cos(m\theta) + \hat{F}_{n,m} \sin(m\theta) + \\ G_{n,m} r^2 \cos(m\theta) + \hat{G}_{n,m} r^2 \sin(m\theta) \end{pmatrix} + \sum_{n=2}^{\infty} r^{-n} \sum_{m=2}^{n} \begin{pmatrix} H_{n,m} \cos(m\theta) + \hat{H}_{n,m} \sin(m\theta) + \\ J_{n,m} r^2 \cos(m\theta) + \hat{J}_{n,m} r^2 \sin(m\theta) \end{pmatrix} \\
\psi_3 &= \sum_{n=2}^{\infty} r^{-n} \sum_{m=2}^{n} \left( C_{n,m} \cos(m\theta) + \hat{C}_{n,m} \sin(m\theta) + E_{n,m} r^2 \cos(m\theta) + \hat{E}_{n,m} r^2 \sin(m\theta) \right),
\end{aligned} \qquad (A4)$$

where the unknown coefficients at each order of perturbation are obtained by applying the boundary condition as follows

(i) flow field satisfies the no-slip and no-penetration boundary condition at the interface

(ii) tangential stress is continuous at the interface.

The expressions of the coefficients in equations (A2) and (A4) have provided as an extract from the MAPLE files in the supplementary material.

We have employed normal stress boundary condition for obtaining the droplet shape. The coefficients of equation (14) are represented as

$$\begin{aligned}
M_{2,0}^{(Ca_E)} &= H + \{G\cosh(At) + J\sinh(At)\}\exp(Bt), \\
L_{2,0}^{(Ca_E)} &= C + \{F\sinh(At) - C\cosh(At)\}\exp(Bt).
\end{aligned} \qquad (A5)$$

Details of the constant coefficients present in equation (A5) are shown in section A of the supplementary material. On the other hand, $M_{2,1}^{(Ca_E^2)}$, $L_{2,1}^{(Ca_E^2)}$, $M_{4,1}^{(Ca_E^2)}$, and $L_{4,1}^{(Ca_E^2)}$ are obtained by solving the following differential equations

$$A_{21}\left(\frac{\partial L_{2,1}^{(Ca_E^2)}}{\partial t}\right) + B_{21}\left(\frac{\partial M_{2,1}^{(Ca_E^2)}}{\partial t}\right) + C_{21}M_{2,0}^{(Ca_E)} + D_{21}L_{2,0}^{(Ca_E)} + E_{21}L_{2,1}^{(Ca_E^2)} + F_{21} = 0$$

$$H_{21}\left(\frac{\partial L_{2,1}^{(Ca_E^2)}}{\partial t}\right) + I_{21}\left(\frac{\partial M_{2,1}^{(Ca_E^2)}}{\partial t}\right) + J_{21}M_{2,0}^{(Ca_E)} + K_{21}L_{2,0}^{(Ca_E)} + M_{21}M_{2,1}^{(Ca_E^2)} + N_{21} = 0,$$

$$A_{41}\left(\frac{\partial L_{4,1}^{(Ca_E^2)}}{\partial t}\right) + B_{41}\left(\frac{\partial M_{4,1}^{(Ca_E^2)}}{\partial t}\right) + \left\{\begin{array}{l}(C_{41}L_{2,0}^{(Ca_E)} + D_{41}M_{2,0}^{(Ca_E)})\left(\frac{\partial L_{2,0}^{(Ca_E)}}{\partial t}\right) + (E_{41}L_{2,0}^{(Ca_E)} + F_{41}M_{2,0}^{(Ca_E)})\left(\frac{\partial M_{2,0}^{(Ca_E)}}{\partial t}\right) \\ + G_{41}L_{2,0}^{(Ca_E)} + H_{41}M_{2,0}^{(Ca_E)} + I_{41}M_{4,1}^{(Ca_E^2)} + J_{41}\left(M_{2,0}^{(Ca)}\right)^2\end{array}\right\} = 0,$$

$$M_{41}\left(\frac{\partial L_{4,1}^{(Ca_E^2)}}{\partial t}\right) + N_{41}\left(\frac{\partial M_{4,1}^{(Ca_E^2)}}{\partial t}\right) + \left\{\begin{array}{l}(O_{41}L_{2,0}^{(Ca_E)} + P_{41}M_{2,0}^{(Ca_E)})\left(\frac{\partial L_{2,0}^{(Ca_E)}}{\partial t}\right) + (Q_{41}L_{2,0}^{(Ca_E)} + R_{41}M_{2,0}^{(Ca_E)})\left(\frac{\partial M_{2,0}^{(Ca_E)}}{\partial t}\right) \\ + S_{41}L_{2,0}^{(Ca_E)} + T_{41}M_{2,0}^{(Ca_E)} + U_{41}L_{4,1}^{(Ca_E^2)} + V_{41}\left(L_{2,0}^{(Ca)}\right)^2\end{array}\right\} = 0.$$

(A6)

Here, all the coefficients of equation (A6) are functions of $(R, S, \lambda, K)$. The expressions of $M_{2,1}^{(Ca^2_E)}$, $L_{2,1}^{(Ca^2_E)}$, $M_{4,1}^{(Ca^2_E)}$ and $L_{4,1}^{(Ca^2_E)}$ are displayed in section B of the supplementary material for certain typical values of $R, S, \lambda, K$.

**Appendix B. Cahn number ($Cn$) and grid size independence test**

Next, we perform a grid independency study to ensure that our outcomes are independent of the size of grid as shown in figure 10. In the present analysis, the Cahn number is equivalent to grid size. So, the numerical results are also independent of Cahn number.

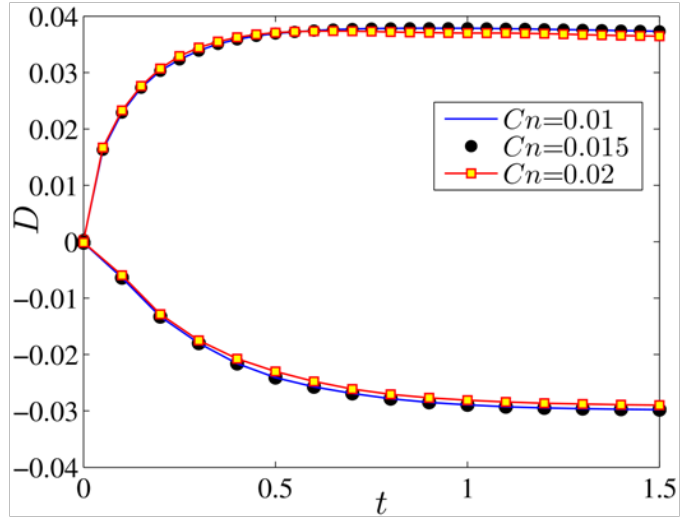

FIGURE 11. Temporal variation of deformation parameter of a LD-LD-LD system having $(S_{23}, R_{23})$=(0.4397, 0.033) and $(S_{23}, R_{23})$=(2.274, 30.33). Other parameters used are $Re = 0.01$, $\lambda=1$ and $Wc=0.2$.

Here, we have shown the transient evolution of the deformation parameter for different values of the Cahn number and it is seen that variation in Cahn number shows a negligible change in the deformation parameter. In the present study, we have chosen $Cn = 0.01$ for all the plots.

**Supplementary material**

The supplementary material contains the expression of the constant coefficients present in equations (A5) and (A6). The expressions of the coefficients in equations (A2) and (A4) are too large to be provided in a doc file. Hence the same has been given as an extract (.pdf file) from the MAPLE files in the supplementary material.

**ACKNOWLEDGMENTS**

S.S. and S.D. are grateful to Dr. Shubhadeep Mandal for insightful discussions on droplet EHD.